# The Statistics of Microlensing Light Curves:
# I. Amplification Probability Distributions

Geraint F. Lewis[1]* and Mike J. Irwin[2]

[1] *Institute of Astronomy, Madingley Rd., Cambridge, CB3 0HA*
[2] *Royal Greenwich Observatory, Madingley Road, Cambridge CB3 0EZ*

12 March 1995

**ABSTRACT**

The passage of stars through the beam of a lensed quasar can induce violent fluctuations in its apparent brightness. The fluctuations observed in the Huchra lens, (2237+0305), are taken to be the first evidence of this "microlensing" occurring in lensing systems. Subsequent microlensing events observed in this system and in other gravitational lenses illustrate that microlensing should be a common phenomenon in lensed quasars. The statistical properties of the component light curves will be defined to a large extent by the mass distribution of the microlensing objects and the internal light distribution of the lensed quasar. We present statistics of a large sample of hypothetical microlensing light curves, for a number of lensing situations, generated using an efficient numerical technique. These artificial light curves show that the amplification probability distributions are independent of the mass function of the compact lensing objects. The amplification probability distributions for the Huchra lens system are presented.

**Key words:** Microlensing, Numerical Simulations

## 1 INTRODUCTION

Einstein's theory of relativity predicts that light will be deflected in the gravitational field of a massive object (Einstein 1915). The "gravitational lensing" due to the mass of the sun was observed by Eddington and Crommlin during a solar eclipse in 1919, confirming the general relativistic theory (Eddington 1919).

For a simple lens component separations are proportional to $M^{1/2}$ where $M$ is the mass of the lens interior to the impact parameter. This implies that at cosmological distances, while galaxy clusters can produce image splittings of several arcseconds, individual galaxies produce arcsecond splittings and stellar masses only micro-arcsecond splittings. Conventionally, gravitational lensing by galaxies and galaxy clusters is referred to as macrolensing, whilst lensing due to individual stars is known as microlensing. At cosmological distances the image splittings due to stellar masses are too small to be directly observable, however their microlensing effect can still induce observable brightness fluctuations in the images of macrolensed objects. Chang and Refsdal (1979) studied the effect of a single star in the beam of a lensed quasar and showed that fluctuations of up to one magnitude on time-scales varying from a few months to several years, dependent on the quasar continuum size, could be produced. However, it was soon realized that in general macrolensing situations the action of multiple compact objects embedded within the smooth surface potential of the macrolens had to be considered. With these higher surface densities of potential microlensing objects the expected variation of brightness with time can no longer be dealt with efficiently using analytic techniques and the generation of realistic microlensing light curves must be handled numerically.

Young (1981) presented the first simulations for an ensemble of stars. He showed that quasars whose beam passes through a starfield should exhibit complex variability in their light curves, although he reveals little about his numerical technique. Other methods dealt mainly with image finding techniques on a grid (Paczyński 1986). These methods were invariably slow (also limited by the then available computing systems) and could not guarantee that all the images, and therefore the full light curve, had been found. With the development of the backward ray-shooting technique [Kayser et al. (1986), Wambsganss (1990)] a map of the amplification as a function of position in the source plane could be readily generated. A light curve is then found by considering the passage of a source across this pattern. As the pattern is two-dimensional, source profiles can be successfully convolved with the amplification map to pro-duce light curves for extended sources. Though powerful the ray-shooting method requires a large number of rays fired through the image plane ( $\sim 10^{12}$ ) and even with the addition of tree code algorithms to speed up the calculations of the deflection angles, many cpu hours are required to generate a single amplification map. This can be very prohibitive in the construction of large samples of independent microlensing light curves. Although several light curves can be drawn from any given map, the ensuing light curves are then not strictly independent.

---

* Email: gfl@mail.ast.cam.ac.uk

In an effort to avoid the shortcomings of the ray shooting method Witt (1993) and independently Lewis et al. (1993) developed an efficient method for directly generating microlensing light curves. This method uses the fact that it is easier to track the image of a straight source line than to find the images of a point source. As a source moves behind a starfield its path can be considered to be such a straight line. The image of the source line consists of an infinite length curve, which tends to a straight line far from the lensing masses, plus a series of closed loops whose loci pass through the microlensing stars. This ensures that the image curves are readily deducible numerically as the starting position of every image curve is known. As these image curves are followed their contribution to the total light curve can be calculated. The method is, therefore, a one dimensional image following routine, rather than two-dimensional ray shooting or image finding, and all the rays intercept the required source track. There is therefore no redundancy in the number of rays fired through the lens plane and fewer rays are required for the generation of each light curve. Light curves can be generated efficiently even at high surface mass densities of stars enabling the statistical properties of a wide variety of different lensing scenarios to be investigated. Several examples of light curves generated using this method are shown in Figure 1. The main disadvantage of this technique is that the two-dimensional convolution required to allow for finite source size has to be done indirectly, a point we discuss in more detail later.

In this paper we present the initial statistics of a large sample of microlensing light curves that were generated using this efficient method. In Section 2 the effect of lensing shear on the images of a microlensed source is discussed. Section 3 discusses the details of the microlensing light curve simulations and Section 4 presents the amplification probability distributions for the sample and discusses the form of these distributions. In Section 5 the distributions for slightly extended sources are presented, while simulations for Huchra Lens appear in Section 6. The conclusions of this study are presented in Section 7.

## 2 THE EFFECT OF SHEAR ON THE IMAGE OF A STRAIGHT LINE

For a smooth matter lens with no external shear the image of a straight line is itself a straight line, oriented parallel to the source. If the smooth matter in the lens is replaced with compact bodies the image of the straight line is seen to deform into loop-like images that start and end on the compact objects, and an infinite, non-linear image [see Witt(1993) and Lewis et al.(1993) for more details]. As these variations are only of order $10^{-6}$ arcseconds on the sky the image of the microlensed line appears unchanged on large scales.

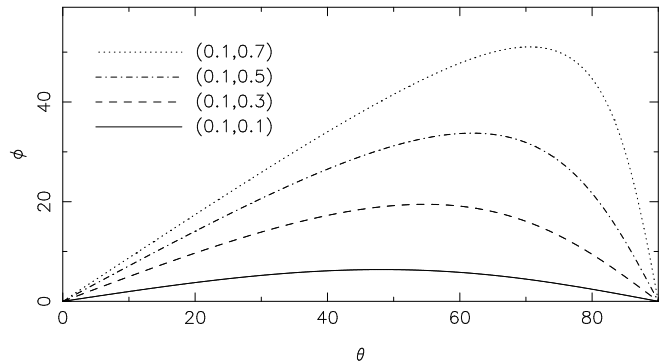

**Figure 2.** The relationship between the orientation of the image of a straight line, $\phi$, and the orientation of the shear, $\theta$ for several lensing scenarios. The numbers in brackets represent $(\sigma_c, \gamma)$, where $\sigma_c$ is the normalised smooth matter density and $\gamma$ is the shear.

With microlensing the shear is a first order effect of the gravitational field, due to the large scale mass distribution, across a bundle of light rays (throughout this paper the term shear refers to this external shear, rather than the shear generated by the point mass lenses). The ellipticity and orientation of extended images, due to shear, have been discussed elsewhere [Schneider, Ehlers & Falco (1992)], but here the effect of this shearing on the image of a straight line is addressed.

Considering a region of a smooth matter lens, which is small compared to the scale of the matter variation, under the influence of a shear, $\gamma$, aligned at an arbitrary angle to the co-ordinate system, the lensing equation is given by

$$\mathbf{y} = \begin{pmatrix} 1 - \gamma_1 & -\gamma_2 \\ -\gamma_2 & 1 + \gamma_1 \end{pmatrix} \mathbf{x} - \sigma_c \mathbf{x}. \qquad (1)$$

Here, $\mathbf{y}$ and $\mathbf{x}$ are angular co-ordinates of the source and images respectively, $\gamma_1 = \gamma\cos(2\theta)$ and $\gamma_2 = \gamma\sin(2\theta)$, with $\theta$ being the direction of the shear with respect to the co-ordinate system. The surface mass density of the smooth matter, $\sigma_c$, is in units of the critical surface density,

$$\sigma_{crit} = \frac{c^2}{4\pi G} \frac{D_{os}}{D_{ol}D_{ls}}, \qquad (2)$$

where $D_{ij}$ are the observer-lens-source angular diameter distances (Schneider et al. 1992).

If the source is an infinite long line running along $y_2 = 0$ the image track is given by the equation

$$x_2 = \frac{\gamma_2}{1 + \gamma_1 - \sigma_c}x_1 \qquad (3)$$

where $(x_1, x_2)$ represents the Cartesian position in the image plane. This equation simply describes a straight line oriented at an angle, $\phi = tan^{-1}(\frac{\gamma_2}{1+\gamma_1-\sigma_c})$, to the co-ordinate system. Figure 2 illustrates the dependence of this angle upon the orientation of the shear angle, $\theta$, for several values of the surface mass density and the shear, $(\sigma_c, \gamma)$.

Figure 3 illustrates the form of the image of a straight

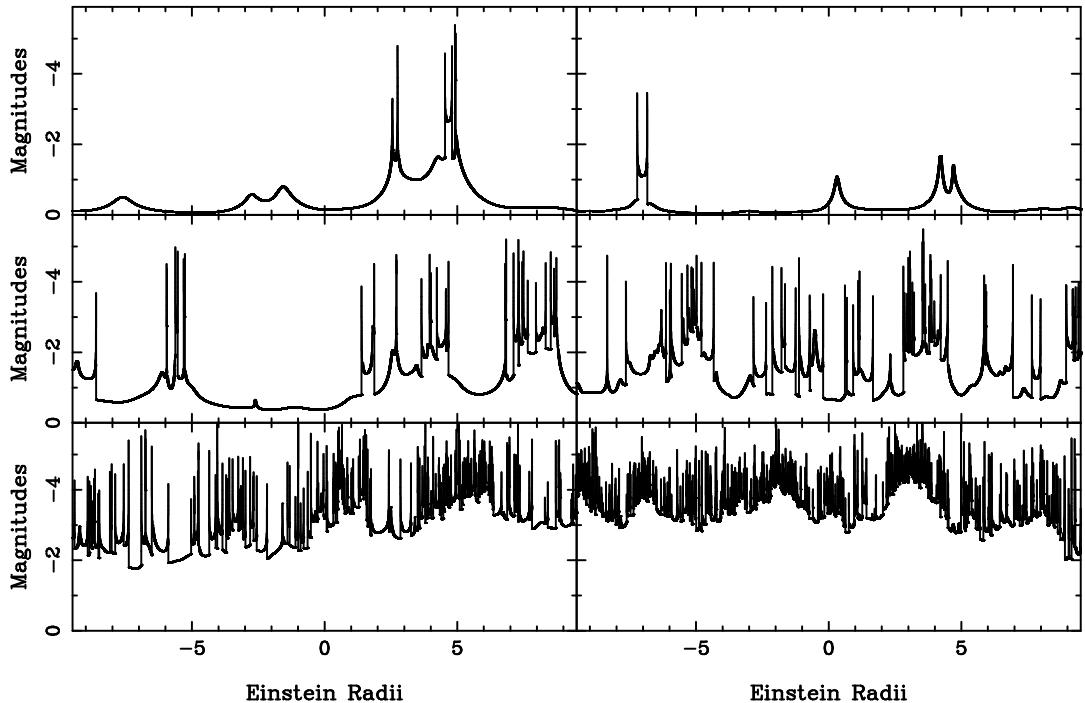

**Figure 1.** Several light curves generated using the method presented in Witt (1993) and Lewis et al.(1993). The microlensing lensing parameters for these simulations are $(\sigma_*, \gamma) =$ (0.2,0.0), (0.5,0.0) and (0.8,0.0), from top to bottom resp.. The light curves on the right are for cases where the lensing objects are distributed with a Salpeter mass distribution between 0.3 and 1.0 $M_\odot$, while those on the left have all the stars at one solar mass (see Section 3)

line when observed through a field of compact object with shear across them, the lensing parameters being $(\sigma_*, \gamma) =$ (0.3, 0.3) with the shear oriented at angle of 30°. Here $\sigma_*$ is the surface mass density of compact object in units of $\sigma_{crit}$. The points represent the positions of the lensing masses, and the solid line represents the lensed image. The dotted line represents the image of the line that would be seen if all the mass in the compact objects were smoothly distributed. The resulting track and images are simply perturbations about the expected result for the smooth matter lens, with additional images existing in the form of loops.

## 3 THE SIMULATIONS

The light curve samples were generated using the efficient method outlined previously. The natural angular scale in a lensing system is the Einstein radius, which is defined as

$$\Theta_E^2 = \frac{4GM}{c^2} \frac{D_{ls}}{D_{os} D_{ol}}, \tag{4}$$

where $M$ is the characteristic lensing mass. With microlensing this mass is taken to be one solar mass. Physically, $\Theta_E$ is the angular radius of the circular image formed when a source at distance $D_{os}$ is perfectly aligned with a point mass lens at distance $D_{ol}$ and the source crossing time of an Einstein radius defines the lensing time scale. To provide a large coverage and hence a better statistical sample each light curve was followed over the equivalent of twenty Einstein radii. Fifty such light curves were generated for each

| Parameters | | MSOLAR | | SALPETER | |
|---|---|---|---|---|---|
| $\sigma_*$ | $\mu_{th}$ | $N_m$ | $<\mu_m>$ | $N_s$ | $<\mu_s>$ |
| 0.20 | 1.56 | 94 | 1.47 | 290 | 1.51 |
| 0.35 | 2.37 | 335 | 2.33 | 940 | 2.43 |
| 0.50 | 4.00 | 989 | 3.86 | 2621 | 4.06 |
| 0.65 | 8.16 | 3045 | 8.18 | 7787 | 8.00 |
| 0.80 | 25.0 | 12966 | 24.8 | 17541 | 24.4 |

**Table 1.** Parameters for the cases with no shear. $\sigma_*$ is the normalized surface density of stars and $N_i$ is the number of stars used in the generation of each light curve. $\mu_{th}$ is the theoretical amplification and $<\mu_i>$ is the mean amplification from the generated light curves.

set of lensing parameters, with several examples presented in Figure 1. The samples generated were:

### 3.1 Samples with no shear

The initial sample consists of star fields with no shear. This reduces the problem to a two parameter model, the surface mass density and the stellar mass function. Two mass functions were used, the first ( **MSOLAR** ) has all the stars of the same mass, namely one solar mass. The second sample ( **SALPETER** ) consists of stars distributed with a Salpeter mass function. This is defined as

$$f(m)dm \propto m^\beta dm, \tag{5}$$

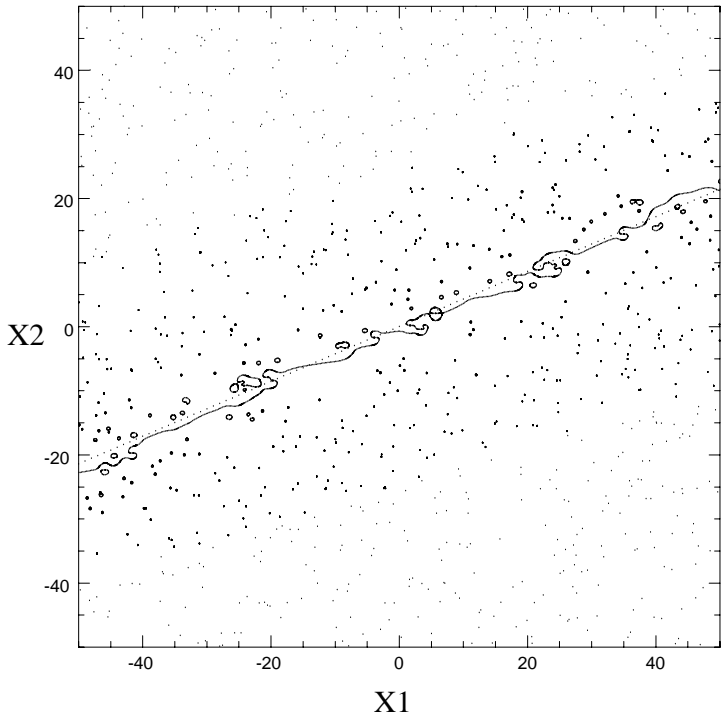

**Figure 3.** The solid curve and loops in this figure represents image of a straight line seen through a star field with external shear. The points represent the positions of the lensing masses, while the dotted line represents the image of the straight line for a smooth matter lens. The lensing parameters are $(\sigma, \gamma) = (0.3, 0.3)$, with the masses of the stars being one solar mass. The shear is oriented at $30°$, and all distances are in units of the Einstein radius (Eqn 4).

with $\beta = -2.35$ between the limits of 0.1 and 1.0 solar masses. Five different values of $\sigma_*$ were used, 0.2, 0.35, 0.5, 0.65 and 0.8, with 50 light curves generated for each different combination of parameters. Details, including the mean amplification of each sample of light curves, are presented in Table 1. The theoretically expected mean amplification for the equivalent smooth mass distribution is given by

$$\mu_{th} = \frac{1}{|(1-\sigma_*)^2 - \gamma^2|}. \qquad (6)$$

Table 1 also presents the number of stars used in each simulation. The number of stars was chosen to ensure that $> 99\%$ of all image light is collected for each point on the light curve. These stars are randomly distributed within a circle of the appropriate radius (Lewis et al. 1993).

### 3.2 Samples with shear

The addition of shear adds two more lensing parameters, the magnitude of the shear and its orientation with respect to the orientation of the source track. A grid of points in the $(\sigma_*, \gamma)$ plane is used to provide coverage of the typical range of microlensing parameters. Table 2 presents the number of stars used in each simulation. For each $(\sigma_*, \gamma)$ six numbers are given. The left hand column presents the number of stars required for the **MSOLAR** case, while the right hand column presents the **SALPETER** case. The top value in each column presents the number of stars required for a shear orientation of $0°$, the middle value for $45°$ and the bottom value for $90°$. The differing values are due to the differing length of the image track, for a source track of constant length, as the shear orientation varies. The $\infty$ elements of this plane are at points where the theoretical amplification is infinite. It is unphysical (and impossible due to the diverging number of stars) to attempt to make microlensing simulations at these points.

Fifty light curves per parameter set, $(\sigma_*, \gamma, \phi, MF)$, were generated. This resulted in a total of 6000 light curves. Table 3 presents the mean amplifications from these samples. For each $(\sigma_*, \gamma)$ three numbers are given. The top number presents the mean for the **MSOLAR** sample, while the middle number is is for the **SALPETER** mass function. These values are average of the means from each shear orientations and the standard deviation, $\sigma_{(n-1)}$, of these results is also given. The lower number is the theoretically expected mean value, given by Equation 6. There is a good agreement between the theoretically expected amplifications and the means of the samples.

## 4 THE AMPLIFICATION PROBABILITY DISTRIBUTIONS

### 4.1 The distributions

The amplification distributions for the cases with no shear are presented as the dotted curves in Figure 4. The value for the surface mass density of stars, $\sigma_*$, is listed on the right hand side of the panels, with the left hand panels presenting the distributions for the **MSOLAR** cases and the right hand panels are those for the **SALPETER** cases. The abscissa is in units of magnitudes with respect to the theoretically expected mean amplification (Eq. 6). The ordinate is in units of $\mathrm{Log}_{10}(p(m))$, where $p(m)$ is the probability of being amplified to between m and m + $\delta$m, where $m = 2.5\log_{10}(\mu/\mu_{th})$, and $\mu$ is the lensing amplification.

| $\gamma\backslash\sigma_*$ | 0.20 | | 0.35 | | 0.50 | | 0.65 | | 0.80 | |
|---|---|---|---|---|---|---|---|---|---|---|
| 0.20 | 173<br>140<br>103 | 519<br>401<br>272 | 706<br>573<br>416 | 1961<br>1500<br>974 | 2748<br>2208<br>1517 | 7274<br>5445<br>3227 | 16583<br>13093<br>8045 | 42409<br>30824<br>15183 | $\infty$ | |
| 0.35 | 304<br>227<br>131 | 924<br>642<br>311 | 1585<br>1201<br>678 | 4403<br>3076<br>1386 | 10994<br>8384<br>4361 | 29122<br>20316<br>7838 | $\infty$ | | 23101<br>18514<br>11800 | 57432<br>42420<br>21877 |
| 0.50 | 667<br>483<br>217 | 2066<br>1358<br>455 | 6371<br>4693<br>2073 | 17670<br>11885<br>3656 | $\infty$ | | 16586<br>12912<br>7145 | 42404<br>30228<br>12608 | 5772<br>4775<br>3474 | 14353<br>11079<br>7017 |
| 0.65 | 2787<br>1952<br>682 | 8329<br>5338<br>1213 | $\infty$ | | 10977<br>8329<br>4073 | 29105<br>20164<br>7042 | 4140<br>3285<br>2075 | 10599<br>7757<br>3989 | 2561<br>2175<br>1073 | 6378<br>5107<br>3610 |
| 0.80 | $\infty$ | | 6336<br>4656<br>1967 | 17699<br>11864<br>3394 | 2742<br>2107<br>1173 | 7269<br>5125<br>2203 | 1834<br>1484<br>1021 | 4701<br>3535<br>2072 | 1443<br>1251<br>1025 | 3583<br>2952<br>2233 |

**Table 2.** Number of stars for the simulations with shear. For each parameter set, $(\sigma_*,\gamma)$, six numbers are given. The left hand column is for cases where all the stars are of one solar mass, while the right hand column is for the cases where the stars are distributed with a Salpeter mass function. The top row is for a shear orientation of $0°$, the middle row for shear of $45°$, and the lowest row is for a shear orientation of $90°$.

| $\gamma\backslash\sigma_*$ | 0.20 | 0.35 | 0.50 | 0.65 | 0.80 |
|---|---|---|---|---|---|
| 0.20 | $1.59\pm 0.05$<br>$1.65\pm 0.03$<br>1.66 | $2.55\pm 0.06$<br>$2.67\pm 0.05$<br>2.61 | $4.77\pm 0.11$<br>$4.71\pm 0.08$<br>4.76 | $12.07\pm 0.58$<br>$11.83\pm 0.34$<br>12.12 | $\infty$ |
| 0.35 | $1.82\pm 0.05$<br>$1.92\pm 0.03$<br>1.93 | $3.22\pm 0.11$<br>$3.27\pm 0.02$<br>3.33 | $7.93\pm 0.34$<br>$7.83\pm 0.02$<br>7.84 | $\infty$ | $12.13\pm 0.32$<br>$11.93\pm 0.54$<br>12.12 |
| 0.50 | $2.42\pm 0.11$<br>$2.59\pm 0.09$<br>2.56 | $5.79\pm 0.12$<br>$5.74\pm 0.05$<br>5.80 | $\infty$ | $7.88\pm 0.24$<br>$7.60\pm 0.08$<br>7.84 | $4.76\pm 0.11$<br>$4.82\pm 0.12$<br>4.76 |
| 0.65 | $4.54\pm 0.02$<br>$4.53\pm 0.15$<br>4.60 | $\infty$ | $5.93\pm 0.25$<br>$5.71\pm 0.23$<br>5.80 | $3.36\pm 0.01$<br>$3.31\pm 0.10$<br>3.33 | $2.58\pm 0.10$<br>$2.58\pm 0.04$<br>2.61 |
| 0.80 | $\infty$ | $4.71\pm 0.10$<br>$4.54\pm 0.16$<br>4.60 | $2.58\pm 0.02$<br>$2.58\pm 0.01$<br>2.56 | $1.86\pm 0.09$<br>$1.96\pm 0.02$<br>1.93 | $1.61\pm 0.05$<br>$1.74\pm 0.09$<br>1.67 |

**Table 3.** Mean amplifications for the simulations with shear. The top entry is for the case of a solar mass function, the middle for a Salpeter mass function. These values are averaged over the three shear orientations at each $(\sigma_*,\gamma)$. The standard deviation of this measure, $\sigma_{(n-1)}$, is also presented. The bottom entry is the theoretically expected value.

The distribution for the cases with the inclusion of shear are presented in Figure 5 for the **MSOLAR** case and Figure 6 for the **SALPETER** case.

The panels are arranged in a similar format to the statistics presented in Table 2, with $\sigma_*$ running horizontally and $\gamma$ running vertically.

Each frame contains three distribution, one for each of the shear orientations listed in Table 2. Again, the distributions are re-normalized to the theoretical mean amplification (Eq. 6).

### 4.2 Discussion

Several of the distributions presented here can be compared directly to those presented in Wambsganss (1992). As the source is finite in the Wambsganss simulations the probability functions taper off at high amplification due to the inherent convolution a finite source represents. The point-like nature of the source used in the method presented here implies

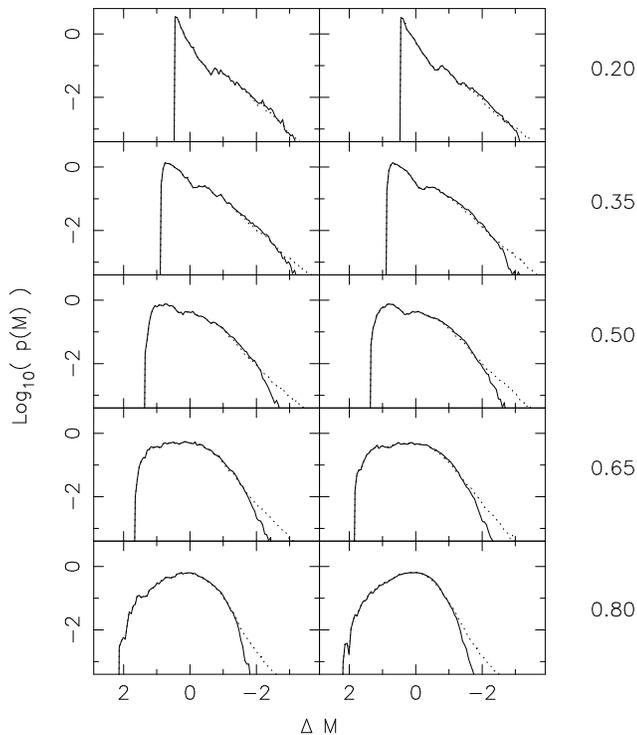

**Figure 4.** The amplification probability distributions for the cases with no shear. The x-axis of each plot is magnitude with respect to the theoretical mean. The left-hand row of panels present the distributions for the Salpeter mass case while the right-hand set of panels are for the case where all the lenses have the same mass. The dotted line represents the amplification probability distribution for the point-like source, while the solid line is for a source of finite size. (Section 5)

that probability distributions drawn from such light curves should possess an $\mu^{-3}$ at high amplification [see Schneider et al. (1992) for details of this tail]. This high amplification tail is seen in all the distributions presented and is illustrated in Figure 7. The amplification probability is for the Salpeter mass case, with $\sigma_* = 0.65$ and no shear. The dashed line illustrates a $\mu^{-3}$ slope and has been displaced vertically for clarity.

Several important features can be seen. Firstly, as expected, the shape of the probability distribution is independent of the orientation of the shear. More importantly there is no dependence of the form of the amplification probability distribution, even at low amplifications, on the mass function of the lensing bodies at any values of $\sigma_*$ and $\gamma$. As an illustration of this further samples of light curves where generated for the case where $\sigma_* = 0.5$ and $\gamma = 0$ and

  (i) all the stars are of mass 0.3 $M_\odot$,
  (ii) all the stars are of mass 10.0 $M_\odot$,
  (iii) the stars are distributed with a Salpeter mass function between the limits of $0.3 M_\odot < M < 10.0 M_\odot$.

The details of these samples are presented in Table 4. The amplification probability distributions for the above mass functions are plotted in Figure 8, as well as the $\sigma_* = 0.5$ cases presented in Section 3.1. The form of the distributions is the same for each of the mass functions indicating that the amplification probability distribution is insensitive to the mass function of the lensing objects.

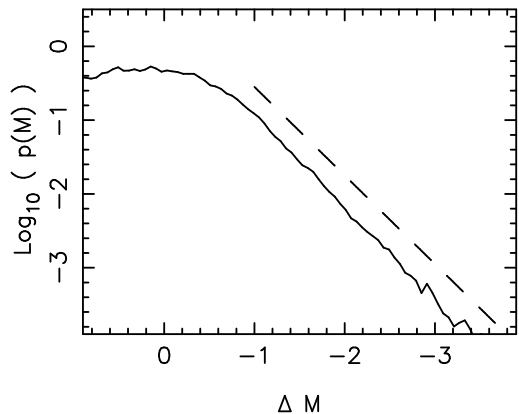

**Figure 7.** An example of the asymptotic behavior of the amplification probability distribution for point sources. The distribution is that of the Salpeter mass case with $\sigma_* = 0.65$ and no shear. The dashed line is the asymptotic $\mu^{-3}$ line displaced vertically for clarity.

| Sample | Mass Function | $N_*$ | $<\mu_{th}>$ |
|---|---|---|---|
| (i) | $m = 0.3\ M_\odot$ | 75659 | 3.95 |
| (ii) | $m = 10\ M_\odot$ | 3837 | 3.92 |
| (iii) | $0.3 < m/M_\odot < 10$ | 33345 | 3.98 |

**Table 4.** The number of stars and the mean amplification of the additional microlensing light curves presented. As $\sigma_* = 0.5$ and $\gamma = 0$ the theoretical mean amplification for these samples is 4.

The theoretical study of the amplification probability distributions of Schneider(1987) showed that the shape of the high amplification tail of the distribution should only depend on $\sigma_*$ and $\gamma$ and not on the mass spectrum. This result could not be extended directly to the complete amplification probability distribution since in general there is no direction link between the amplification probability distribution of an individual microimage and the amplification probability distribution of the many microimages making up the image of a source. The result we have found numerically extends Schneider's conjecture to cover the complete probability distribution function. The detailed shape of the amplitude probability distribution is independent of the mass spectrum of the lensing objects and depends only on the value of the external shear and the value of the surface mass density. The corollary of this result is that measurement of the amplification probability distribution, or properties derived from it, reveal nothing about the mass distribution of the compact objects.

It is apparent from Figure 5 and Figure 6 that the form of the amplification probability distribution is strongly dependent upon $\sigma_*$ and $\gamma$. The secondary "bump" seen in many of the distributions was discussed in Rauch et al. (1992). Their analysis of low optical depth and no shear lensing cases revealed secondary bumps in the probability distributions at moderate amplifications. These features are not immediately apparent in the amplification probability distributions at higher $(\sigma_*, \gamma)$ published in Wambsganss (1992), although the distributions published here exhibit a strong double peak structure over a large region of the parameter space. Rauch et al. interpreted these features as secondary probability distributions due to the creation of pairs of very bright images at caustics. A prediction of this study is that higher order bumps may be visible in the probability distributions at high amplifications, though at a very low level. These are not seen in any of the distributions presented in this work.

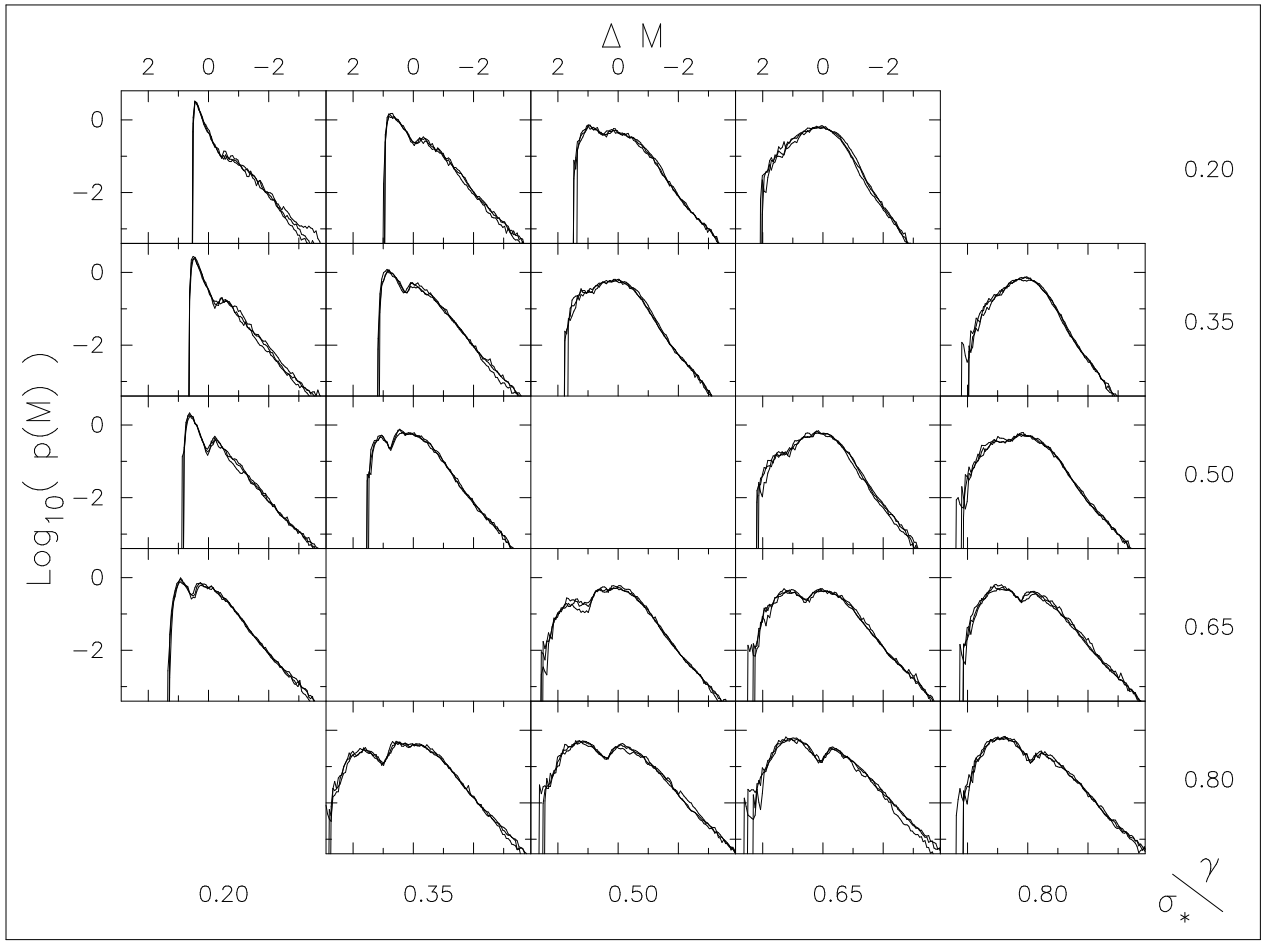

**Figure 5.** The amplification probability distributions for the cases with shear and all the stars of one solar mass. The x-axis of each plot is magnitude with respect to the theoretical mean while the y-axis is $Log_{10}(p(m))$. Each panel contains distributions for the three shear angles $(0°, 45°, 90°)$.

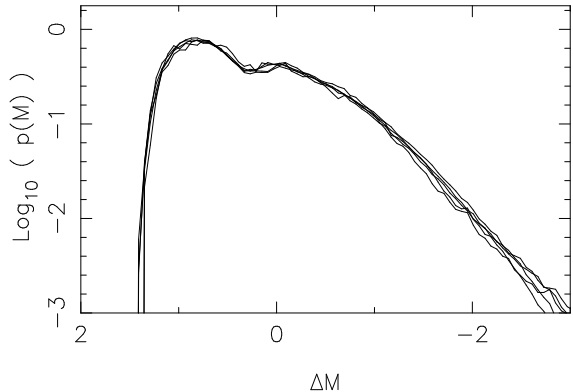

**Figure 8.** The amplification probability distributions for the various mass functions at $\sigma_* = 0.5$. The amplification probability distributions are statistically identical for the various simulations illustrating that the entire form of the distribution is independent of the mass function of the lensing objects.

Although we will discuss the detailed temporal properties of the microlensing light curves in a later paper, one of the approaches we have adopted leads to a useful statistic of the amplification probability distribution and we include it here for completeness. In traditional time series analysis global statistical properties of the signal are derived using some variant of the signal correlation function. In the case of real microlensing light curves, the discrete irregular sampling of the observations is well suited for correlation analysis based on the signal temporal structure functions (Rutman 1978). In particular the first order structure function is closely related to the signal autocorrelation function and is straightforward to compute. For a discrete series of magnitudes, $m_i, i = 1,2....n$, a robust first order structure function can be defined as

$$\psi_j = <| m_{i+j} - m_i |> \qquad (7)$$

where the average $<>$ is taken over all available values and $j$ is the lag between samples [eg. Hook et al. (1994)]. In the asymptotic limit, as the lag $j$ tends to large values, the structure function defined above provides a single measure of the intrinsic variability of the lensed source, namely the average difference in magnitudes between widely separated points. We can compute this quantity directly from the amplitude probability distributions, p(m), by noting that the probability distribution of the difference in magnitudes is simply the convolution of p(m) with p(-m), and hence $\psi_j$ in the limit of large $j$, directly measures the spread or width of p(m). Figure 9 shows how this width varies for the values of $(\sigma_*, \gamma)$ presented in this work. Each curve represents a differing value of the shear, $\gamma$, as indicated in the box. There is strong evolution of this statistic with both $\sigma_*$ and $\gamma$, with high values of $\gamma$ and medium values of $\sigma_*$ providing the strongest signal. Fitting the structure function, Eqn. 7, with a functional form allows the characterization of the microlensing variability via a time-scale. The asymptotic value deduced from the amplification probability distribution provides a normalization of this functional form.

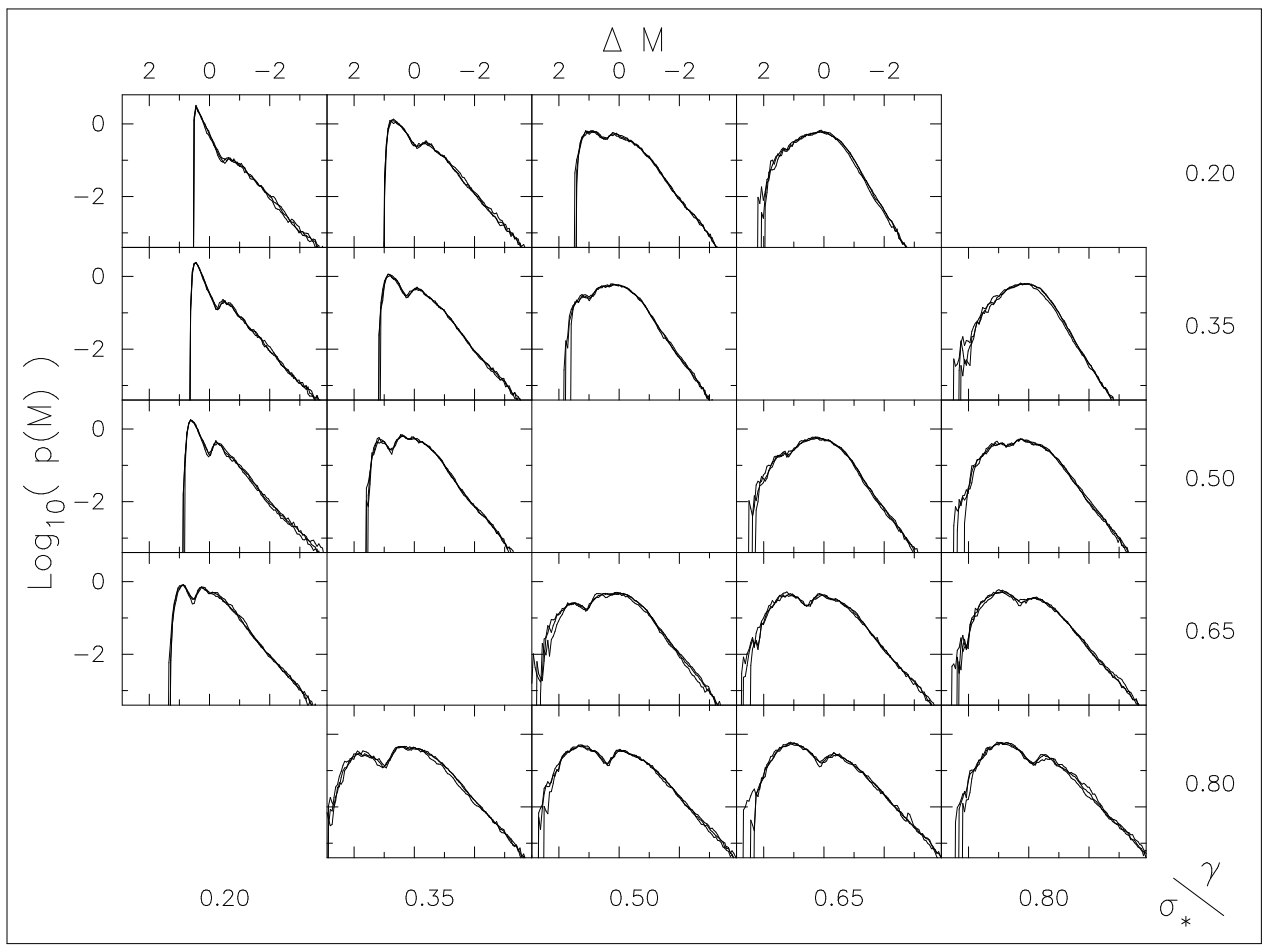

**Figure 6.** As for Figure 5 but for the Salpeter mass case

For the distributions presented here it is clear that with the inclusion of shear the secondary "bump" dominates the shape (and width) of the overall distribution. Therefore, the shape (and width) of observed amplification probability distributions can provide a useful probe of the macrolensing parameters of a system. There is little point in further parameterising the form of these probability distributions since simple pattern recognition techniques can be used to match the numerically derived distributions with any observed equivalents. Furthermore, by pattern matching directly, allowance can be made for noise in the observations, the effects of irregular finite sampling and so on. This would be difficult to achieve using parameterised forms.

## 5 THE EFFECT OF AN EXTENDED SOURCE

The amplification of an extended source can be found by convolving the source profile with the two-dimensional amplification map in the source plane. The method of Witt (1993) and Lewis et al. (1993) is purely one-dimensional in nature and the form of the amplification tangential to the source track is unknown. Witt [in Surdej et al. (1993)] showed that, if the source is small compared to the scale of an Einstein radius, the two dimensional source profile can be convolved with a one dimensional microlensing light curve to produce a light curve for an extended source. His light curves were compared to those generated with the ray shooting technique and showed good agreement. This technique breaks down if the source size is much larger than a fraction of an Einstein radius.

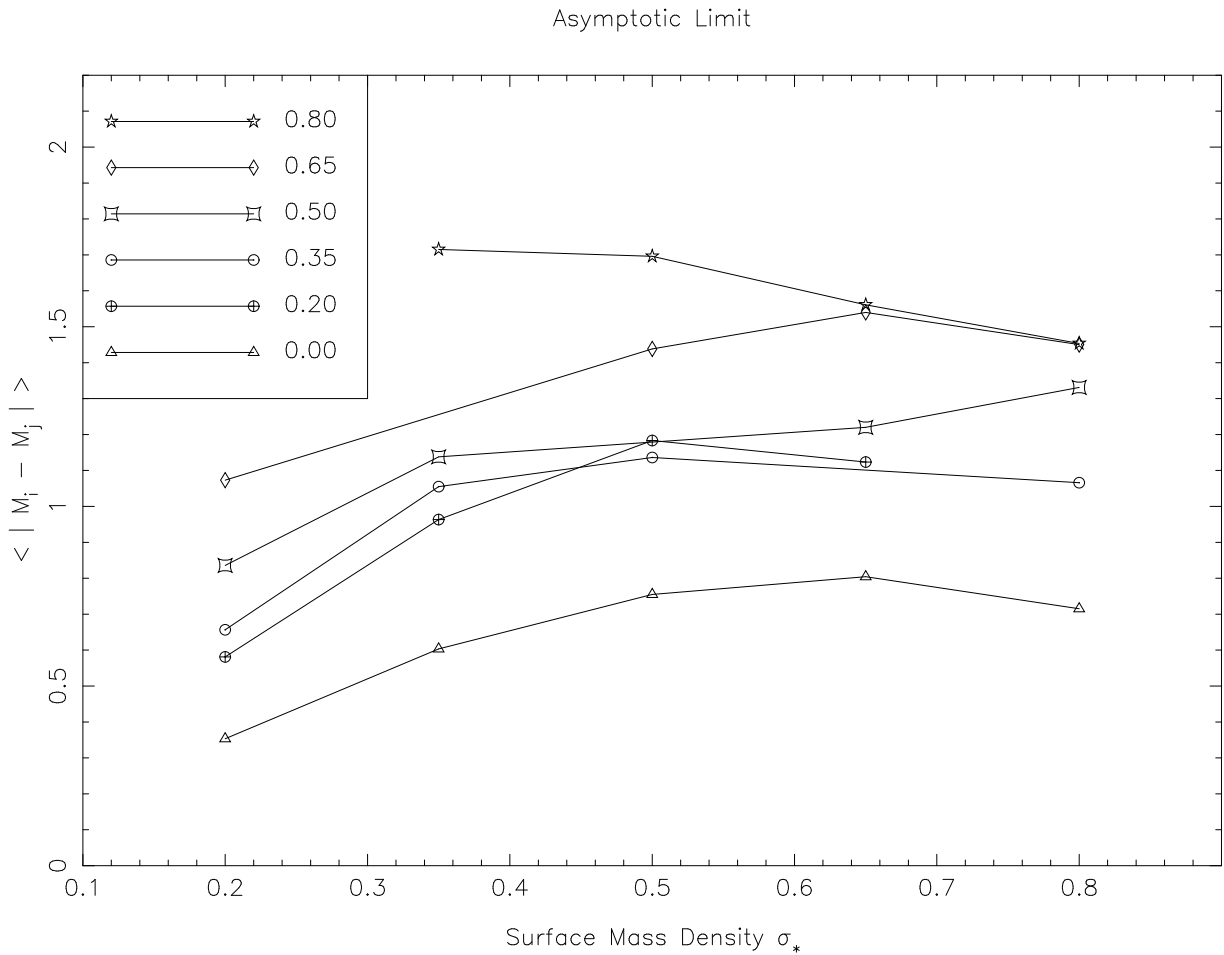

**Figure 9.** Asymptotic limits for the $< |M_i - M_j| >$ statistic, drawn from the amplification probability distributions presented in this paper. The x-axis presents the surface mass density in microlensing objects, $\sigma_*$, while the asymptotic limit is presented on the y-axis. Each curve is for a fixed shear, $\gamma$, whose key can be found in the inset box.

The effect of finite source size was examined for the samples with no shear. The source was chosen to be a uniform disk of radius 0.01 Einstein radii, corresponding to $\sim 1 \times 10^{15} cm$ for the Huchra lens. This corresponds to the expected continuum source size for a typical quasar powered by a supermassive black hole (Rees 1984). We have experimented with other shapes for the source profile but find no significant difference for all reasonable source profiles. The only significant parameter is the effective smoothing scale size of the source. Alternative quasar continuum source models require source sizes significantly large than an Einstein radius and as mentioned earlier are impractical to simulate using our one-dimensional method. The probability distributions for the smoothed light curves are presented in Figure 4. As expected, the probability distributions cut off at the high amplification end. At lower $\sigma_*$ the source extension has little effect on the form of the amplification probability distribution, but at higher values the high end $\mu^{-3}$ tail is noticeably cut away. These distributions resemble those for extended sources, generated with the ray-shooting method, presented in Wambsganss (1992).

## 6 AMPLIFICATION PROBABILITY DISTRIBUTIONS FOR THE HUCHRA LENS

Monitoring of the Huchra Lens (Huchra et al. 1985), a system of four images of a z=1.695 quasar observed through the bulge of a low redshift galaxy, has shown that the component images vary in brightness. This variation is uncorrelated between the images, particularly a rapid brightening and dimming of Image A [using the notation of Yee (1988)] and has been interpreted as microlensing due to stars in the lensing galaxy (Irwin et al. 1989; Corrigan et al. 1991).

Further monitoring has revealed other variations in Images A and B, while Images C and D have remained relatively quiescent exhibiting only long term changes (Corrigan et al. 1991; Racine 1992).

The variation of image brightness in the 2237+0305 system is taken as the first evidence of cosmological microlensing, though recent monitoring of other lensing systems, such as 0957+561 (Thomson and Schild 1995) and 1208+1011 (Hjorth et al. 1995), has revealed similar, uncorrelated variation in image brightness. These observations suggests that microlensing is a common phenomenon in lensing systems.

Modeling of the macrolensing of this system (Kent and Falco 1988; Kochanek 1991) has produced varying estimates for the values of $\sigma$ and $\gamma$ at the positions of the images. Assuming a constant mass:to:light ratio the underlying observed galaxy light distribution has been used to generate a surface mass density distribution enabling $\sigma$ and $\gamma$ to be inferred using the known image positions (and mean amplifications even though these are poorly known) as constraints (Schneider et al. 1988). This technique was employed by Rix using recent HST observations (Rix et al. 1992). The parameters for the best fitting model, consisting of an $R^{\frac{1}{4}}$ elliptical mass distribution with an unresolved core [model 2a in Rix(1992)], are presented in Table 5. [One should note, however, that Kochanek (1991), Witt et al. (1994) and Wambsganss and Paczyński (1994) have shown that any reasonable mass distribution with elliptical symmetry can be used to fit the image positions.]

Samples of fifty light curves were generated for each of these parameter sets, for the case of all the stars being of $1M_\odot$ and a Salpeter mass function between 0.1 and $1M_\odot$. Again, separate orientations of the source trajectory $(0°, 45°, 90°)$ to the shear were undertaken. Details of mean amplifications drawn from the light curve samples are presented in Table 5. These values are averaged over the three shear orientations and the standard deviation of these results, $\sigma_{(n-1)}$, are also given. Table 6 presents the number of stars used in the generation of each light curve. Each light curve was generated to be twenty Einstein radii in length. For the Huchra lensing system a microlens travelling at a typical stellar velocity will cross an Einstein radius in $\approx 7$ years (Wambsganss 1990). This implies that each light curve is $\approx 140$ years in length.

The amplification probability distributions for the images in the Huchra lens are presented in Figure 10. All the distributions are quite broad, showing a double peak structure. The lensing parameters for these images in this system put them quite close to the diagonal in the $\sigma_* - \gamma$ plane where the distributions show this evolved form. The probability distributions for images C and D are quite broad compared to those for images A and B, with more probability of being deamplified. The macroparameters for the D image predict that the image should be brighter than image B, but since its discovery image D has been consistently more than half a magnitude fainter than image B (Lewis et al. 1995). This may be explained by the higher probability of image D being deamplified, though time-scale arguments from a study of the caustic structure by Witt (1993) suggest that, even with the sparse sampling of current observations, some variation should have been seen image D.

| Image | $\sigma_*$ | $\gamma$ | $\mu_{\rm th}$ | $<\mu_{\rm m}>$ | $<\mu_{\rm s}>$ |
|---|---|---|---|---|---|
| A | 0.41 | 0.47 | 7.86 | $7.81 \pm 0.25$ | $7.76 \pm 0.13$ |
| B | 0.38 | 0.43 | 5.01 | $5.04 \pm 0.05$ | $5.01 \pm 0.05$ |
| C | 0.65 | 0.68 | 2.94 | $2.88 \pm 0.13$ | $2.99 \pm 0.08$ |
| D | 0.59 | 0.56 | 6.87 | $6.94 \pm 0.06$ | $6.98 \pm 0.22$ |

**Table 5.** Macrolensing parameters for the best solutions to the lens in 2237+0305 (Rix et al. 1992). $<\mu_{\rm m}>$ and $<\mu_{\rm s}>$ are the mean amplifications of the **MSOLAR** and the **SALPETER** samples resp. This value is averaged over the three shear directions. $\mu_{\rm th}$ is the theoretically expected average.

| Image | 0° | 45° | 90° | 0° | 45° | 90° |
|---|---|---|---|---|---|---|
| A | 12949 | 9427 | 4248 | 34409 | 23397 | 7301 |
| B | 4496 | 3356 | 1639 | 12332 | 8420 | 3015 |
| C | 3419 | 2723 | 1754 | 8760 | 6446 | 3419 |
| D | 14224 | 10963 | 5776 | 35854 | 25975 | 10086 |

**Table 6.** Number of stars in the simulations of the Huchra Lens. The angle refers to the orientation of the shear to the source trajectory. The first three orientations are for a mass function where all the stars are of one solar mass, the second three are for the Salpeter mass distributions.

## 7 CONCLUSIONS

Microlensing may prove to be an important contributor to the variability of lensed quasars, and the statistics of this variability may shed light on the nature of the compact object within lensing galaxies. The mass function of the compact objects could reveal the presence of dark matter in the form of brown dwarfs and could also test theories of star formation. This paper has presented initial statistics of a large sample of independent light curves, generated with an efficient method, over a large volume in macrolensing parameter space.

Schneider (1987) showed that the high-end tail of the amplification probability distribution due to microlensing should be insensitive to the mass functions of the compact objects. The distributions presented here confirm this result over a large volume of parameter space, even at low amplifications where Schneider's theoretical analysis did not apply. The implication of this result is that properties drawn from these probability distributions will not probe the mass spectrum of lensing bodies. However, these statistics will be sensitive to the values of the macrolensing parameters due to the strong evolution of the form of this distribution.

Light curves for extended sources can be achieved via the convolution of the source brightness profile with the one dimensional light curves generated in this paper. This implies that statistical properties drawn from large samples of "realistic" light curves would differentiate between microlensing parameter for observed lensing systems, such as 2237+0305 (given sufficient monitoring).

Visual inspection of light curves with the same macrolensing parameters but differing mass functions reveals that the distribution of time-scales of microlensing are sensitive to the mass function and the relative direction of the shear across the starfield. The statistical properties of the temporal variation of the lensed light curves of events and their dependence upon the mass function is the subject of further work.

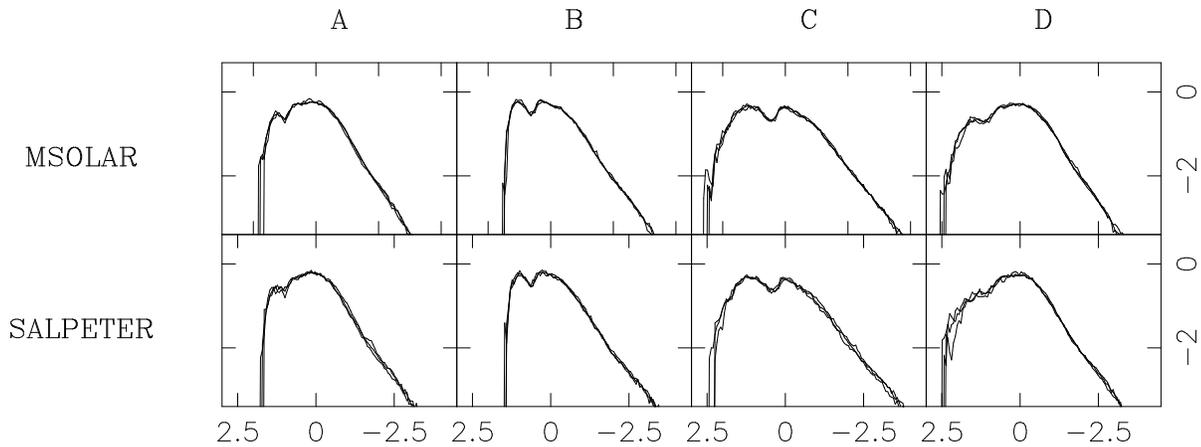

**Figure 10.** The amplification probability distributions for the four images in the Huchra lens. The top set of panels are for cases where all the stars are of one solar mass and the lower panels are for the Salpeter mass case. All the distributions are normalised to the theoretically expected mean.

## 8 ACKNOWLEDGMENTS

Jordi Miralda-Escudé is thanked for valuable discussions. GFL would like to thank all the users of the Cambridge Astronomical Sun Fleet for continued support during this study. We also thank the anonymous referee for useful comments.

## REFERENCES

Chang K., Refsdal S., 1979 Nature, 282, 561
Corrigan R. T. et al., 1991, AJ, 102, 34
Eddington A. S., 1919 Observatory, 42, 119
Einstein A., 1915 *Sitzungber. Preuβ. Akad. Wissensch., erster Halbband*, p. 831
Hjorth J., Grundahl F., Nilsson K., Festin L., 1995, preprint
Hook I. M., McMahon R. G., Boyle B. J., Irwin M. J., 1994, MNRAS, 268, 305
Huchra J., Gorenstein M., Kent S, Shapiro I., Smith G., Horine E., Perley R., 1985, ApJ, 90, 691
Irwin M. J., Webster R. L., Hewett P. C., Corrigan R. T., Jedrzejewski R. I., 1989, AJ, 98, 1989
Kayser R., Refsdal S., Stabell R., 1986, AA, 166, 36
Kent S. M., Falco E. E., 1988, AJ, 96, 1570
Kochanek C. S., 1991, ApJ, 373, 354
Lewis G. F., Miralda-Escudé J., Richardson D. C., Wambsganss J., 1993, MNRAS, 261, 247
Lewis G. F., Irwin M. J., Hewett P. C., 1995 In preparation
Paczyński B., 1986, ApJ, 301, 503
Racine R., 1992, AJ, 102, 454
Rauch K. P., Mao S., Wambsganss J., Paczyński B., 1992, ApJ, 386, 30
Rees M. J., 1984, Ann. Rev. Ast. and Ast., 22, 471
Rix H. W., Schneider, D. P., Bahcall, J. N., 1992, ApJ, 104, 959
Rutman J., 1978, Proc. IEEE, 66, 1048
Schneider D. P., Turner E. L., Gunn J. E., Hewitt J. N., Schmidt M., Lawrence C. R., 1988, AJ, 95, 1619
Schneider P., 1987, ApJ, 319, 9
Schneider P., Ehlers J., Falco E. E., 1992, Gravitational Lensing. Springer–Verlag, Berlin.
Surdej J,. Fraipont-Caro D., Gosset E., Refsdal S., Remy M., 1993 Gravitational Lenses in the Universe. Université de Liège, Belgique
Thomson D. J., Schild R., 1995 preprint
Wambsganss J., 1990, "Gravitational Microlensing" Report MPA, Garching
Wambsganss J., 1992, ApJ, 386, 19
Wambsganss J., Paczyński B., 1994, AJ, 108, 1156
Witt H. J., 1993, ApJ, 403, 530
Witt H. J., Kayser R., Refsdal S., 1993, AA, 268, 501
Witt H. J., Mao S., Schechter P., 1995, preprint
Yee H. K. C., 1988, AJ, 95, 1331
Young P., 1981 ApJ, 244, 756